\documentclass[prd,twocolumn,showpacs,twoside,preprintnumbers,amsmath,amssymb,floatfix,nofootinbib,superscriptaddress]{revtex4-1}
\usepackage{amsmath,amssymb,bm,graphicx,bbold,epsf,colordvi}
\usepackage{lipsum}
\usepackage{braket}
\usepackage{epsfig}
\usepackage{dcolumn}
\usepackage{bbold}
\usepackage{wasysym}
\usepackage{hyperref}
\usepackage{color,soul}
\allowdisplaybreaks 
\def \be  {\begin{equation}}
\def \ee  {\end{equation}}
\def \ba  {\begin{eqnarray}}
\def \ea  {\end{eqnarray}}
\def \baa {\begin{eqnarray*}}
\def \eaa {\end{eqnarray*}}
\def \nn {\nonumber}

\newcommand\f{\frac}

\begin{document}
\title{Jet angularities in photoproduction at the Electron-Ion Collider}
\author{Elke-Caroline Aschenauer}
\email{elke@bnl.gov}
\affiliation{Physics Department, Brookhaven National Laboratory, Upton, NY 11973, USA}
\author{Kyle Lee}
\email{kunsu.lee@stonybrook.edu}
\affiliation{C.N. Yang Institute for Theoretical Physics, Stony Brook University, Stony Brook, NY 11794, USA}
\affiliation{Department of Physics and Astronomy, Stony Brook University, Stony Brook, NY 11794, USA}
\author{B.S. Page}
\email{bpage@bnl.gov}
\affiliation{Physics Department, Brookhaven National Laboratory, Upton, NY 11973, USA}
\author{Felix Ringer}
\email{fmringer@berkeley.edu}
\affiliation{Physics Department, University of California, Berkeley, CA 94720, USA}
\affiliation{Nuclear Science Division, Lawrence Berkeley National Laboratory, Berkeley, CA 94720, USA}

\begin{abstract}
We consider the one-parameter family of jet substructure observables known as angularities using the specific case of inclusive jets arising from photoproduction events at an Electron-Ion Collider (EIC). We perform numerical calculations at next-to-leading logarithmic accuracy within perturbative QCD and compare our results to PYTHIA~6 predictions. Overall, we find good agreement and conclude that jet substructure observables are feasible at the EIC despite the relatively low jet transverse momentum and particle multiplicities. We investigate the size of subleading power corrections relevant at low energies within the Monte Carlo setup. In order to establish the validity of the Monte Carlo tune, we also perform comparisons to jet shape data at HERA. We further discuss detector requirements necessary for angularity measurements at an EIC, focusing on hadron calorimeter energy and spatial resolutions. Possible applications of precision jet substructure measurements at the EIC include the tuning of Monte Carlo event generators, the extraction of nonperturbative parameters and studies of cold nuclear matter effects.
\end{abstract}

\maketitle

\section{Introduction}

A high energy and high luminosity Electron-Ion Collider (EIC) will be the ideal machine to explore in detail the structure of nucleons and nuclei. Jet observables are expected to play a major role in this effort, complementing measurements of identified hadrons in the final state. Recently, various inclusive jet measurements and correlations have been proposed at the EIC~\cite{Kang:2012zr,Kang:2013nha,Hinderer:2015hra,Abelof:2016pby,Hinderer:2017ntk,Chu:2017mnm,Klasen:2017kwb,Boughezal:2018azh,Klasen:2018gtb,Gutierrez-Reyes:2018qez,Liu:2018trl,Gutierrez-Reyes:2019vbx,Gutierrez-Reyes:2019msa,Hatta:2019ixj,Zheng:2018awe,Dumitru:2018kuw}. An advantage of jets is that they can be calculated purely perturbatively whereas hadron inclusive cross sections require knowledge of nonperturbative fragmentation functions. In addition, jet measurements extend the kinematic range compared to observables involving hadrons and can provide unique constraints on collinear PDFs, transverse momentum dependent PDFs and fragmentation functions (FFs). The measurement of different processes is important to assess universality aspects of QCD factorization. An identified jet also allows for a clean separation of the current and target fragmentation region. Moreover, jets are also a useful tool to better understand cold nuclear matter effects in $e+A$ collisions. Other related work on physics opportunities at an EIC can be found in~\cite{Boer:2011fh,Accardi:2012qut,Aschenauer:2017oxs}.

In this work, we systematically explore for the first time the feasibility of jet substructure measurements at the EIC. The substructure of jets has gained increased attention in the past years at RHIC and the LHC both in proton-proton and heavy-ion collisions. Corresponding measurements in the relatively clean environment at the EIC can provide important complementary information. Some of the recent advancements of jet substructure tools can be applied directly at the EIC where jets allow for precision tests of QCD in $e+p$ and $e+A$. Possible applications include the tagging of quarks and gluons, tagging of the initial state, spin correlations, the measurement of fragmentation functions, studies of hadronization, tuning of parton showers and extractions of the strong coupling constant of QCD.

Jets at the EIC will have relatively small transverse momenta $p_T$ and low particle multiplicities~\cite{Aschenauer:2017jsk}. Nevertheless, as we demonstrate in this work, jet substructure measurements are feasible at a future EIC. We highlight challenges both from the theoretical and experimental sides and we find that some definitions of jet substructure observables are better suited for EIC physics than others. As a first example, we consider in this work jet angularities~\cite{Berger:2003iw,Almeida:2008yp,Ellis:2010rwa,Larkoski:2014pca}, which are defined as
\begin{equation}\label{eq:taua}
    \tau_a=\frac{1}{p_T}\sum_{i\in J}p_{Ti}\Delta R_{iJ}^{2-a} \,.
\end{equation}
Here $p_{Ti}$ are the transverse momenta of the particles relative to the beam axis and $\Delta R_{iJ}$ is their distance to the jet axis in the $\eta$-$\phi$ plane. The sum over all particles inside the jet $i\in J$ is normalized by the total jet transverse momentum $p_T$. Jet angularities assign a single number $\tau_a$ to the identified jet characterizing its radiation pattern. The parameter $a$ smoothly interpolates between traditional jet substructure observables such as the jet mass $(a=0)$ and jet broadening $(a=1)$. 

In this work we present results for jet angularities for EIC kinematics obtained within perturbative QCD. The obtained results are compared to parton shower simulations obtained within a Monte Carlo (MC)-framework based on PYTHIA~6~\cite{Sjostrand:2006za}. This comparison also allows us to assess nonperturbative aspects of jet substructure observables at the EIC. To ensure the validity of the MC-framework in this regime we compare to jet substructure data from HERA and find a good agreement. At electron-proton colliders jets can be measured in different frames and we have to classify events into low photon virtuality $Q^2$, quasi-real photoproduction\footnote{Throughout the rest of this work, the quasi-real production of photons is simply referred to as photoproduction.}, and high virtuality, Deep Inelastic Scattering (DIS), or we can choose not to observe the final state electron. In this work we are generally interested in the feasibility of jet substructure observables at the EIC and without loss of generality we choose to work in the laboratory/center of mass (CM) frame and we consider the quasi-real photoproduction cross section of jets. Jet studies in photoproduction processes can be particularly useful in order to constrain the elusive parton-in-photon distribution functions~\cite{Chu:2017mnm}. We specify the detector requirements needed to perform jet substructure measurements at the EIC both in $e+p$ and $e+A$ collisions.

The remainder of this paper is organized as follows. 
In Section~\ref{sec:theory}, we discuss the relevant perturbative QCD (pQCD) factorization framework that we employ to make predictions for jet angularities at the EIC. In Section~\ref{sec:pythia}, we review the the MC framework and in section~\ref{sec:numerics}, we present comparisons of our results obtained within pQCD and PYTHIA for inclusive jets and jet angularities. EIC detector requirements to perform jet substructure measurements are discussed in section~\ref{sec:detector}. In section~\ref{sec:conclusions}, we conclude and present an outlook.
 
\section{QCD factorization framework \label{sec:theory}}

In this section, we introduce the QCD factorization formalism for jet angularities at the EIC. We first consider the cross section for inclusive jet production and then extend the formalism to include the jet angularity measurement performed on the constituents of the observed jet.

\subsection{The quasi-real photoproduction of inclusive jets}

\begin{figure}[t]
\vspace*{.7cm}
\includegraphics[width=8.5cm]{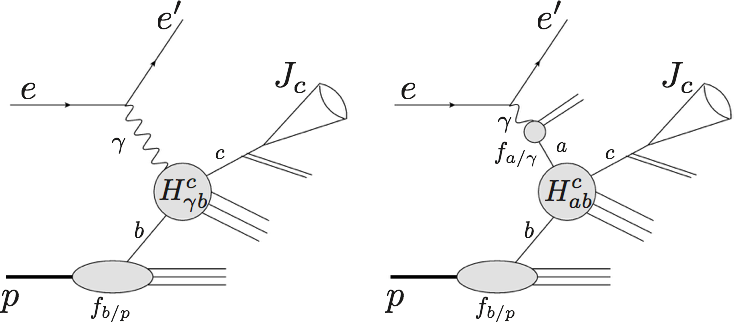}
\caption{The direct (left) and resolved (right) contribution to photoproduction cross section of inclusive jets $e + p\to e' + {\rm jet}+X$ at the EIC.~\label{fig:direct_resolved}}
\end{figure}

We consider the photoproduction cross section of inclusive jets $e + p\to e'+{\rm jet}+X$ at small values of the photon virtuality $Q^2$ at the EIC. We work in the laboratory or center of mass frame. We note that at electron-hadron colliders, jets have often been measured in the Breit frame. However, for example the jet shape/jet energy profile was measured at HERA in the laboratory frame~\cite{Breitweg:1998gf}. Also in~\cite{Hinderer:2015hra,Abelof:2016pby,Hinderer:2017ntk,Boughezal:2018azh,Liu:2018trl} the jet cross section was considered in the laboratory frame. One of the advantages of this choice is that it allows for a more direct comparison to jets in proton-proton collisions. The photoproduction cross section can be separated into a direct and a resolved contribution
\begin{equation}
    {\rm d}\sigma={\rm d}\sigma_{\rm dir}+{\rm d}\sigma_{\rm res} \,.
\end{equation}
The two cases are illustrated in Fig.~\ref{fig:direct_resolved}. The nearly on-shell photon can interact with the partons directly, or the photon can resolve into its parton content which requires us to introduce the nonperturbative parton-in-photon PDFs. Using QCD factorization, we can then write both contributions differential in the jet transverse momentum $p_T$ and the center of mass (CM) frame rapidity $\eta$ as
\begin{eqnarray}\label{eq:incl_factorization}
\frac{{\rm d}\sigma^{e p\to e'\, {\rm jet}+X}}{{\rm d}Q^2\, {\rm d}p_T\, {\rm d}\eta}    &=&\sum_{abc} f_{a/e}(x_e,\mu)\otimes f_{b/p}(x_p,\mu)\nonumber\\
&&\hspace*{-1.5cm}\otimes\, H_{ab}^c(x_a,x_b,p_T/z,\eta,\mu)\otimes J_c(z,p_T R,\mu)\,.
\end{eqnarray}
Note that we included here $Q^2$ on the left hand side to differentiate this case from the $Q^2$ integrated case which was considered in~\cite{Hinderer:2015hra,Abelof:2016pby} which includes both photoproduction and the large-$Q^2$ DIS regime. In practice, the photon virtuality is integrated to some upper cutoff $Q^2<Q_{\rm{max}}^2$ which is determined by the experimental setup. Note that in the case of photoproduction and the $Q^2$ integrated case, the hard scale is solely set by the jet $p_T$ making the perturbative expansion of the cross section in powers of the strong coupling constant feasible. In DIS there are two perturbative hard scales $Q^2$ and $p_T$. In Eq.~(\ref{eq:incl_factorization}), $f_{b/p}(x_p,\mu)$ denotes the PDF to find parton $b$ in the proton with momentum fraction $x_p$ at the scale $\mu$. In addition, we introduce the effective PDF for finding parton $a$ in the electron carrying a momentum fraction $x_e$ from the electron. For the direct contribution as shown on the left side of Fig.~\ref{fig:direct_resolved}, we have $a=\gamma$. For the resolved contribution, right side of Fig.~\ref{fig:direct_resolved}, the parton $a=q,g$ is obtained from the resolved photon. We can write $f_{a/e}$ as
\begin{equation}\label{eq:leptonphoton}
f_{a/e}(x_e,\mu)=\int_{x_e}^1\frac{{\rm d}y}{y} P_{\gamma e}(y)\, f_{a/\gamma}\Big(x_{\gamma}=\frac{x_e}{y},\mu\Big)
\end{equation}
with the Weizs\"acker-Williams photon spectrum~\cite{deFlorian:1999ge,Frixione:1993yw}
\begin{eqnarray}
P_{\gamma e}(y)&=&\frac{\alpha}{2\pi}\left[\frac{1+(1-y)^2}{y}\ln\frac{Q_{\rm max}^2(1-y)}{m_e^2 y^2} \right.\nonumber\\
&&\left. + \,2 m_e^2 y \left(\frac{1}{Q_{\rm max}^2}-\frac{1-y}{m_e^2 y^2}\right) \right]
\end{eqnarray}
where $\alpha$ is the QED fine structure constant and $m_e$ is the electron mass. For the direct contribution we have simply
\begin{equation}
    f_{a/\gamma}(x_\gamma,\mu)=\delta(1-x_\gamma) \,,
\end{equation}
in Eq.~(\ref{eq:leptonphoton}). Instead, for the resolved contribution we need parton-in-photon PDFs which constitute an additional nonperturbative input. For the calculation within perturbative QCD we use the GRS99 (Gluck, Reya, Schienbein) set of parton-in-photon PDFs of~\cite{Gluck:1999ub} throughout this work. The PDFs of~\cite{Gluck:1999ub} were extracted at NLO using the DIS$_{\gamma}$ scheme. However, they can be converted to the conventional $\overline{\rm MS}$ scheme which we use in this work~\cite{Bardeen:1978yd,Gluck:1999ub}. Other fits of parton-in-photon PDFs can be found in~\cite{Gordon:1991tk,Gluck:1994tv,Schuler:1995fk,Gordon:1996pm,Gluck:1991ee}. The hard functions $H_{ab}^c$ for the scattering process $ab\to c$ in Eq.~(\ref{eq:incl_factorization}) for the resolved contribution are the same as for hadroproduction in proton-proton collisions $p + p\to h+X$. They were calculated analytically to NLO in~\cite{Aversa:1988vb,Jager:2002xm}. The direct contribution was obtained analytically in~\cite{Aurenche:1986ff,Gordon:1994wu,deFlorian:1998fq,Jager:2003vy}. Finally, $J_c$ are the semi-inclusive jet functions calculated in~\cite{Kaufmann:2015hma,Kang:2016mcy,Dai:2016hzf,Kang:2017mda}. Similar to parton-to-hadron fragmentation functions, they take into account the formation of a jet with radius $R$ which is initiated by an active parton $c$. The jet carries a longitudinal momentum fraction $z=p_T/\hat p_T$ of the initial fragmenting parton $c$. The semi-inclusive jet functions satisfy DGLAP evolution equations which allow for the resummation of single logarithms of the jet radius parameter $\alpha_s^n\ln^n R^2$~\cite{Catani:2013oma,Dasgupta:2014yra,Kang:2016mcy,Dai:2016hzf}
\begin{equation}\label{eq:dglap}
    \mu\frac{{\rm d}}{{\rm d} \mu}J_i=\frac{\alpha_s}{2\pi}\sum_{j}P_{ji}\otimes J_j\,.
\end{equation}
Here the $P_{ji}(z)$ are the Altarelli-Parisi splitting functions which are the same as for fragmentation functions. We note that formally Eq.~(\ref{eq:incl_factorization}) holds up to power corrections of order ${\cal O}(R^2)$. It was found that these power corrections are typically small even for large values of the jet radius, both for inclusive jet cross sections as well as jet substructure observables~\cite{Mukherjee:2012uz,Kang:2018qra}. The jet functions also have the advantage that one may directly calculate quark/gluon fractions beyond leading-order, and they allow for a convenient calculation of jet angularities at the EIC as discussed in the next section. The same factorization structure holds for inclusive hadron cross sections where the jet functions in Eq.~(\ref{eq:incl_factorization}) need to be replaced with nonperturbative parton-to-hadron fragmentation functions. The photoproduction of hadrons at the EIC was first considered in~\cite{Jager:2003vy,deFlorian:2013taa}. 

\subsection{Jet angularities}

In order to study jet angularities, we consider the following multi-differential cross section
\begin{equation}
\frac{1}{\sigma_{\rm incl}} \frac{{\rm d}\sigma^{e p\to e'+{\rm jet}+X}}{{\rm d}Q^2\,{\rm d}p_T \, {\rm d}\eta\, {\rm d}\tau_a} \,,
\end{equation}
where $\sigma_{\rm incl}$ denotes the inclusive jet cross section discussed in the previous section. The relevant factorization structure can be obtained from the jet angularity cross section of inclusive jet production in proton-proton collisions~\cite{Ellis:2010rwa,Hornig:2016ahz,Kang:2018qra}. The modification relative to the inclusive cross section in Eq.~(\ref{eq:incl_factorization}) amounts to replacing the semi-inclusive jet function $J_c$ with a jet function that not only depends on the momentum fraction $z$ contained in the observed jet but which also depends on the jet angularity $\tau_a$ of the jet which we denote by ${\cal G}_c(z,p_T R,\tau_a,\mu)$. Following~\cite{Hornig:2016ahz,Kang:2018qra}, we can refactorize the jet function ${\cal G}_c$ in the phenomenologically relevant kinematic limit $\tau_a^{1/(2-a)}\ll R$ in terms of hard-matching functions ${\cal H}_{i\to j}$, collinear $C_j$ and soft functions $S_j$ as
\begin{eqnarray}\label{eq:refactorize}
\hspace*{-.3cm}{\cal G}_c(z,p_T R,\tau_a,\mu)&=&{\cal H}_{i\to j}(z,p_T R,\mu) \nonumber \\
&&\times C_j(\tau_a,p_T,\mu)\otimes S_j(\tau_a,p_T,R,\mu) \,.
\end{eqnarray}
Here $\otimes$ denotes a convolution in the variable $\tau_a$. At the one loop level, the functions ${\cal H}_{i\to j}$ are given by out-of-jet radiation contributions~\cite{Kang:2017mda,Kang:2017glf} which were included before also in the semi-inclusive jet function for inclusive jet production. The factorization here was achieved within Soft Collinear Effective Theory (SCET)~\cite{Bauer:2000ew, Bauer:2000yr, Bauer:2001ct, Bauer:2001yt,Beneke:2002ph} which leads to separate renormalization group (RG) equations for the different functions. The characteristic scales of the three function in Eq.~(\ref{eq:refactorize}) are given by
\begin{equation}
    \mu_{\cal H}\sim p_T R \,,\quad \mu_C\sim p_T\tau_a^{1/(2-a)} \,,\quad \mu_S\sim p_T\tau_a R^{a-1}\,.
\end{equation}
The scale of the hard functions in Eq.~(\ref{eq:incl_factorization}) is given by the hard scale of the process, which is the transverse momentum of the observed jet $\mu_H= p_T$. By solving the associated RG evolution equations, and evolving the functions to a common scale, the all order resummation of large logarithms of the form $\alpha_s^n\ln^{2n}(\tau_a^{1/(2-a)}/R)$ is achieved, which we carry out at next-to-leading logarithmic accuracy (NLL$'$). The fixed order expressions of the involved functions and the relevant anomalous dimensions can be found in~\cite{Kang:2018qra}. Similar to the semi-inclusive jet function $J_c$ in Eq.~(\ref{eq:dglap}), the angularity dependent jet functions satisfy DGLAP evolution equations which allow for the resummation of single logarithms of the jet radius $R$. 

Note that in Eq.~(\ref{eq:refactorize}) only the hard-matching functions ${\cal H}_{i\to j}$ depend on $z$, which is the convolution variable in Eq.~(\ref{eq:incl_factorization}). We can thus completely separate the $z$ dependence from the collinear and soft functions which depend on the jet substructure variable $\tau_a$. This separation allows for the calculation of quark/gluon fractions beyond leading-order in the resummation region. Note that the jet angularities considered here are only nonzero when two partons are inside the jet, i.e. at fixed order ${\cal O}(\alpha_s)$ or overall ${\cal O}(\alpha^2 \alpha_s^2)$. However, in the resummation region we need to take into account out-of-jet radiation diagrams in the matching functions ${\cal H}_{i\to j}$ even though this leaves only one parton in the jet at ${\cal O}(\alpha_s)$. The nonzero value for the jet angularity is then generated by the resummation. This counting in powers of $\alpha_s$, instead of starting from a nonzero value of the jet substructure observable required at fixed order, allows for perturbatively calculable quark/gluon fractions beyond leading-order. The main difference of photoproduction compared to DIS processes or the $Q^2$ integrated result eventually amounts only to changing the calculated quark/gluon fractions, making the results presented in this work broadly applicable. This includes also scattering cross sections with longitudinally polarized initial states. Note that the relevant hard functions for jet production at large-$Q^2$, can be obtained from~\cite{Kniehl:2004hf,Daleo:2004pn,Wang:2019bvb}. Higher fixed order results for jet production can be found in~\cite{Abelof:2016pby,Currie:2018fgr}.

There are several contributions that are not captured by the factorization of the cross section for jet angularities in Eqs.~(\ref{eq:incl_factorization}) and~(\ref{eq:refactorize}). First there are power corrections ${\cal O}(R^2)$ which include for example Initial State Radiation (ISR). Second, there are hadronization corrections since our calculation is carried out at the parton level whereas the experimental data, and the Monte Carlo results presented below, are at the hadron level. In addition, there is a contribution from the underlying event/multi-parton interactions for the resolved contribution. However, this contribution is expected to be much smaller than for example at proton-proton colliders. Therefore, the EIC will provide a clean environment for jet measurements where the dominant nonperturbative correction are primarily due to hadronization effects. We capture these different nonperturbative effects by a shape function~\cite{Korchemsky:1999kt,Lee:2006nr,Stewart:2014nna}. See also~\cite{Dasgupta:2007wa} for a discussion of nonperturbative effects. When the softest scale in the factorization theorem $\mu_S\sim p_T\tau R^{a-1}$ approaches $\sim \Lambda_{\rm QCD}$, nonperturbative effects become important which, start in the region
\begin{equation}\label{eq:NPregion}
    \tau_a\sim \f{\Lambda_{\rm QCD}}{p_T R^{a-1}} \,.
\end{equation}
The purely perturbative cross section ${\rm d}\sigma^{\rm pert}$ following Eq.~(\ref{eq:refactorize}) is then convolved with a shape function $F(k)$, where $\tau_a$ is shifted by the virtuality of the soft mode as
\begin{eqnarray}\label{eq:shapefct1}
    \f{{\rm d}\sigma}{{\rm d}Q^2\,{\rm d}p_T \, {\rm d}\eta\, {\rm d}\tau_a} & = &\nn \\ 
    &&\hspace*{-2.3cm} \int {\rm d}k\, F(k)\f{{\rm d}\sigma^{\rm pert}}{{\rm d}Q^2\,{\rm d}p_T \, {\rm d}\eta\, {\rm d}\tau_a} \Big(\tau_a-\f{k}{p_T R^{a-1}} \Big) \,.
\end{eqnarray}
Following~\cite{Stewart:2014nna} we use a single parameter shape function which is given by
\begin{equation}\label{eq:shapefct2}
    F(k)=\f{4k}{\Omega_a^2}\exp(-2k/\Omega_a) \,,
\end{equation}
which is normalized to unity since the hadronization is expected to affect only the shape of the distribution but not the normalization. In addition, its first moment is given by $\Omega_a$. We factor out the $a$ dependence of the nonperturbative parameter $\Omega_a$ following the work of~\cite{Lee:2006nr} on angularities in $e^+e^-$ collisions
\begin{equation}
  \Omega_a=\f{\Omega_{a=0}}{1-a} \,.  
\end{equation}
Achieving a better understanding of universality aspects of nonperturbative corrections will be an important goal of jet substructure studies at the EIC. Similar shape functions are needed for jets measured in $e^+e^-$ and proton-proton collisions. For example, they play an important role in extractions of the QCD strong coupling constant~\cite{Abbate:2010xh,Hoang:2015hka}.

Furthermore, there are non-global logarithms (NGLs) for the jet angularities which are not captured by our factorization structure, see for example~\cite{Dasgupta:2001sh,Larkoski:2015zka,Balsiger:2019tne}. NGLs are typically most relevant near the nonperturbative regime. While a more rigorous treatment is desirable in the future, we effectively absorb the NGL contribution in the nonperturbative shape function. This may lead to a parameter of the shape function that is larger than the typical nonperturbative scale.

\begin{figure*}[t]\centering
\vspace*{.7cm}
\includegraphics[width=6.7in]{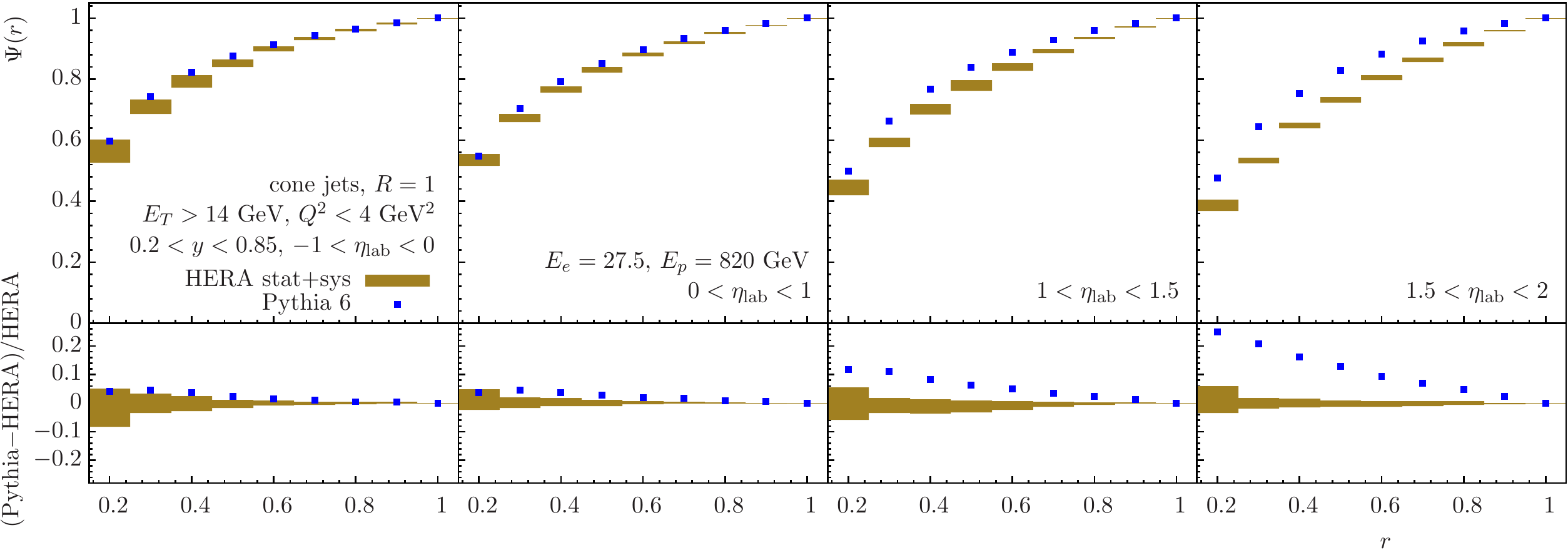}
\caption{Comparison of the integrated jet shape $\Psi(r)$ from HERA measured on an inclusive jet sample in photoproduction events and results from our PYTHIA~6 Monte Carlo. This illustrates the validity of our PYTHIA~6 tune in the context of jet substructure measurements.}
\label{fig:heraComp}
\end{figure*}

One of the concerns of jet substructure observables at relatively low energies is the size of power corrections. In general, there are different types of power corrections that can be important. First, there are perturbative corrections ${\cal O}(\tau_a^2)$. Second, there are corrections which are power suppressed by the large scale of the process, which is here the jet transverse momentum $p_T$. We do not include these corrections in our calculations, but in principle, both are perturbatively calculable. In addition, there are hadronization corrections that we include here using a shape function as discussed above. In order to assess the first type of power correction listed here, we use a different definition of the jet angularity, which agrees with the one in Eq.~(\ref{eq:taua}) up to corrections of order ${\cal O}(\tau_a^2)$. Angularities as an $e^+e^-$ event shape were first introduced in~\cite{Berger:2003iw}. Applied to jet angularities, this definition was given in terms of pseudorapidities $\eta_{iJ}$ and transverse momenta $\vec p_T^{\; iJ}$ relative to the jet axis
\begin{equation}\label{eq:tauap}
    \tau_a'=\frac{1}{2E_J}\sum_{i\in J}|\vec{p}_T^{\; iJ}|\exp(-|\eta_{iJ}|(1-a))\,,
\end{equation}
where $E_J$ is the jet energy, which is not a boost-invariant quantity at hadron colliders. Here we introduced the notation $\tau_a'$ to differentiate between this definition of the jet angularities and the one given in Eq.~(\ref{eq:taua}) above. At small values of the jet angularity in the resummation region, the definition in Eq.~(\ref{eq:tauap}) can be related to the boost invariant definition appropriate for the EIC and hadron collisions given in Eq. (\ref{eq:taua}) up to power corrections
\begin{equation}\label{eq:pc1}
    \tau_a=\left(\frac{2E_J}{p_T}\right)^{2-a}\tau_a' + \mathcal{O}(\tau_a^2)\,.
\end{equation}
For the jet mass case ($a=0$) we can also employ the definition which is often used at hadron colliders
\begin{equation}\label{eq:tauapp}
    \tau_0''= \f{m_J^2}{p_T^2}= \f{1}{p_T^2}\bigg(\sum_{i\in J}p_i \bigg)^2 \,,
\end{equation}
where we sum over all four momenta $p_i$ of the particles inside the observed jet and square the result normalized by $1/p_T^2$. This definition agrees with the other definitions in the small jet mass limit and if the four-momenta of the particles in the jet are taken to be massless
\begin{equation}
    \tau_0=\tau_0''+{\cal O}(\tau_0^2) \,.
\end{equation}
However, this definition is directly sensitive to hadron mass effects ${\cal O}(m_h^2)$. While we do not distinguish between these definitions in our perturbative calculations, we can address how sensitive jet angularities at the EIC are to such power corrections by studying the three different definitions in the Monte Carlo simulations described in the next section.

\section{Monte Carlo setup and validation \label{sec:pythia}}

In principle, it is possible to perform more quantitative comparisons between parton shower event generators and analytical resummations. However, here we are 
interested instead in the feasibility of jet substructure measurements at the future EIC and therefore compare the results obtained within the framework 
detailed in the section above to pseudo-data generated using a Monte Carlo tuned to reproduce HERMES semi-inclusive DIS $e+p$ data over an wide range in $Q^2$, $z$ and $p_T$ \cite{Airapetian:2012ki,Airapetian:2010ac}. The Monte Carlo used is PYTHIA-6 \cite{Sjostrand:2006za} with the CTEQ5m \cite{Lai:1999wy} and SAS 1D-LO \cite{Schuler:1995fk} proton and photon PDFs, respectively.
The older CTEQ5m set was chosen because its PDF 
is not frozen at the input scale $Q_{0}^{2}$ (typically on the order of 1~GeV$^2$) like more modern PDFs and thus returns reliable 
cross sections in the $Q^2 < Q_{0}^{2}$
region addressed in this paper. The SAS PDF was used because it describes the H1 data that is sensitive to photon structure well and, 
as it treats the vector meson and 
anomalous components of the photon wave function separately, avoids double counting issues when simulating subprocesses with 
resolved photons in PYTHIA. Comparisons of the described MC tune with some H1 and ZEUS results and further details on the Monte Carlo can be found in \cite{Zheng:2018awe,Chu:2017mnm}.
Resolved processes, in which the virtual photon interacts via the hadronic component of its wavefunction, dominate 
production of high-$p_{\mathrm{T}}$ particles in the photoproduction region, however, a significant fraction of jets arise from the direct 
processes of photon gluon fusion (PGF) and QCD-Compton (QCDC). These subprocesses are combined and shown separately from the resolved. 
results.

Jets were reconstructed in the laboratory frame from all stable, final-state particles (excluding the scattered beam electron) with 
transverse momenta greater than 250~MeV/$c$ and pseudorapidity of $\pm 4$. Here, stable refers to particles which would normally not
decay within the volume of the detector. Clustering was done using the anti-k$_\mathrm{T}$ algorithm \cite{Cacciari:2008gp} as implemented in the FastJet package \cite{Cacciari:2011ma} with E-scheme recombination  and jet resolution parameters of $R = 0.4$ and 0.8. Further event and jet cuts are listed in Sec.~\ref{sec:numerics}.

The simulation setup used in this manuscript matches that from \cite{Chu:2017mnm}, where it was shown to reproduce HERA dijet cross sections. Further determinations of the suitability of this simulation to describe substructure observables was done by comparing 
to jet shape results from ZEUS in the photoproduction region \cite{Breitweg:1997gg}. The jet shape is defined as
\begin{equation}
\Psi(r) = \frac{1}{N_{\mathrm{J}}}\sum_{\mathrm{J}}\frac{E_{\mathrm{T}}(r)}{E_{\mathrm{T}}(r=R)},
\end{equation}
\noindent where $N_{\mathrm{J}}$ is the total number of jets and $E_{\mathrm{T}}(r)$ is the amount of transverse energy contained within a cone 
of radius $r$ (which is less than or equal to the jet radius $R$) centered on the jet axis. The only modification to the simulation was to 
match the beam energies with HERA ($E_e = 27.5$~GeV and $E_p = 820$~GeV) and the jet-finding was done using a version of the CDF Midpoint Cone algorithm with $R=1$ as 
implemented in FastJet to more closely match the ZEUS analysis \cite{Breitweg:1997gg}. Particle level jets were found from all simulated events 
with $Q^2 < 4$~GeV$^2$ and $0.2 < y < 0.85$ and were required to have transverse energies greater than 14~GeV. 

\begin{figure}[t]\centering
\vspace*{.7cm}
\includegraphics[width=3.5in]{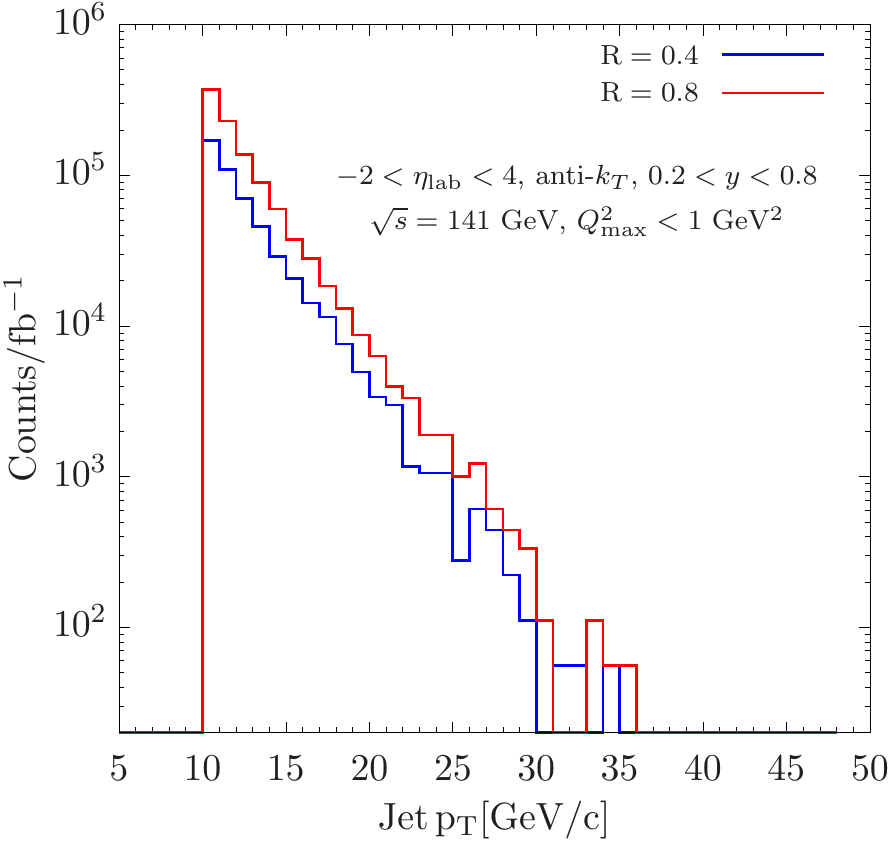}
\caption{The expected inclusive jet yield as a function of jet transverse momentum from photoproducion events at the EIC as returned by our Monte Carlo for 1~fb$^{-1}$ of integrated luminosity. We show the results for $R=0.8$ (red) and $R=0.4$ (blue).~\label{fig:jetmul}}
\end{figure}

Figure \ref{fig:heraComp} presents the comparison between the ZEUS data and our simulation for four jet pseudorapidity ranges. The uncertainty bands  
represent a quadrature sum of statistical and systematic uncertainties on the ZEUS data. It is seen that our simulation reproduces the ZEUS results well for $\eta_{\rm lab} < 1.5$ with 
moderate deviations appearing in the $1.5 < \eta_{\rm lab} < 2.0$ region. This could be due to a number of factors, including the limited modeling of the underlying event and the precision of the photon PDFs. Having improved MC models and more differential data from EIC will help to disentangle these different contributions.
In all cases, the simulation produces more collimated jets than what is observed in data.
It should be noted that the same behavior was seen with the original ZEUS simulations. The relatively good agreement seen between data and simulation give 
confidence that our Monte Carlo can produce in-jet energy distributions close to what will be seen at an EIC.

\section{Numerical results \label{sec:numerics}}

\begin{figure}[t]\centering
\vspace*{.7cm}
\includegraphics[width=3.5in]{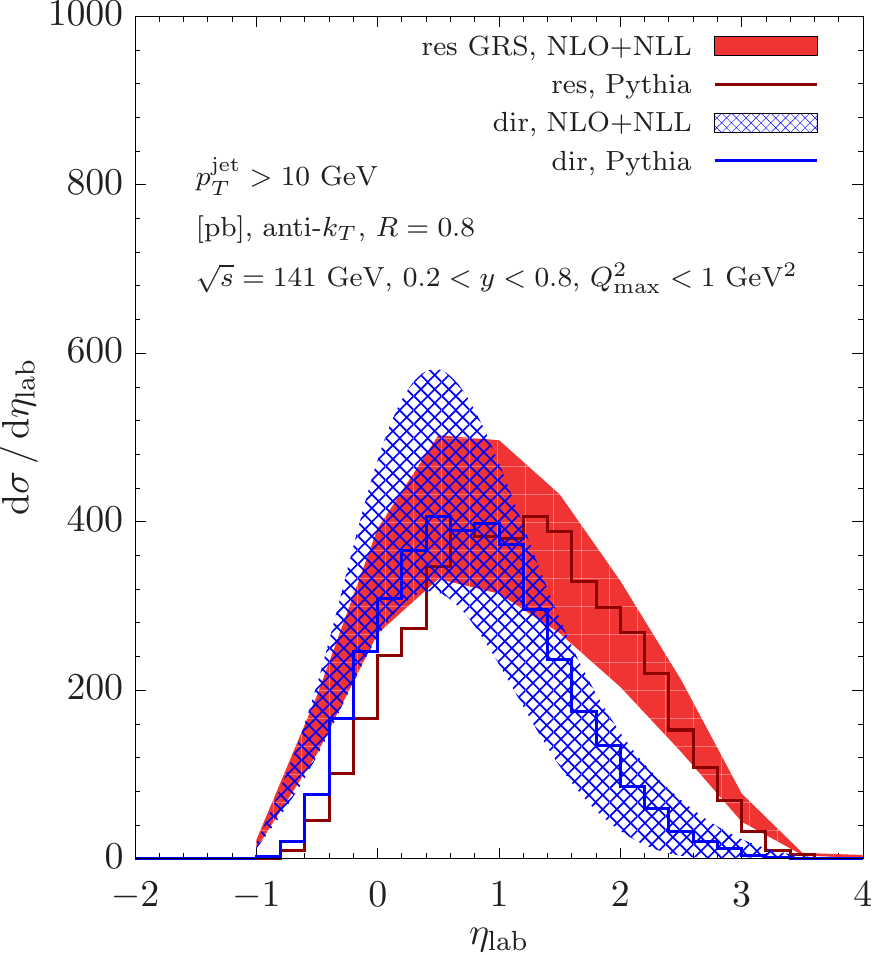}
\caption{The inclusive production cross section of jets in photoproduction events at the EIC as a function of the rapidity in the laboratory frame $\eta_{\rm lab}$. We show the perturbative QCD result for the resolved and direct contribution (red and blue bands) as well as the Monte Carlo result (dark red and blue histograms). The relevant kinematics are displayed in the figure.~\label{fig:incl}}
\end{figure}

\begin{figure*}[t]\centering
\vspace*{.7cm}
\includegraphics[width=5in]{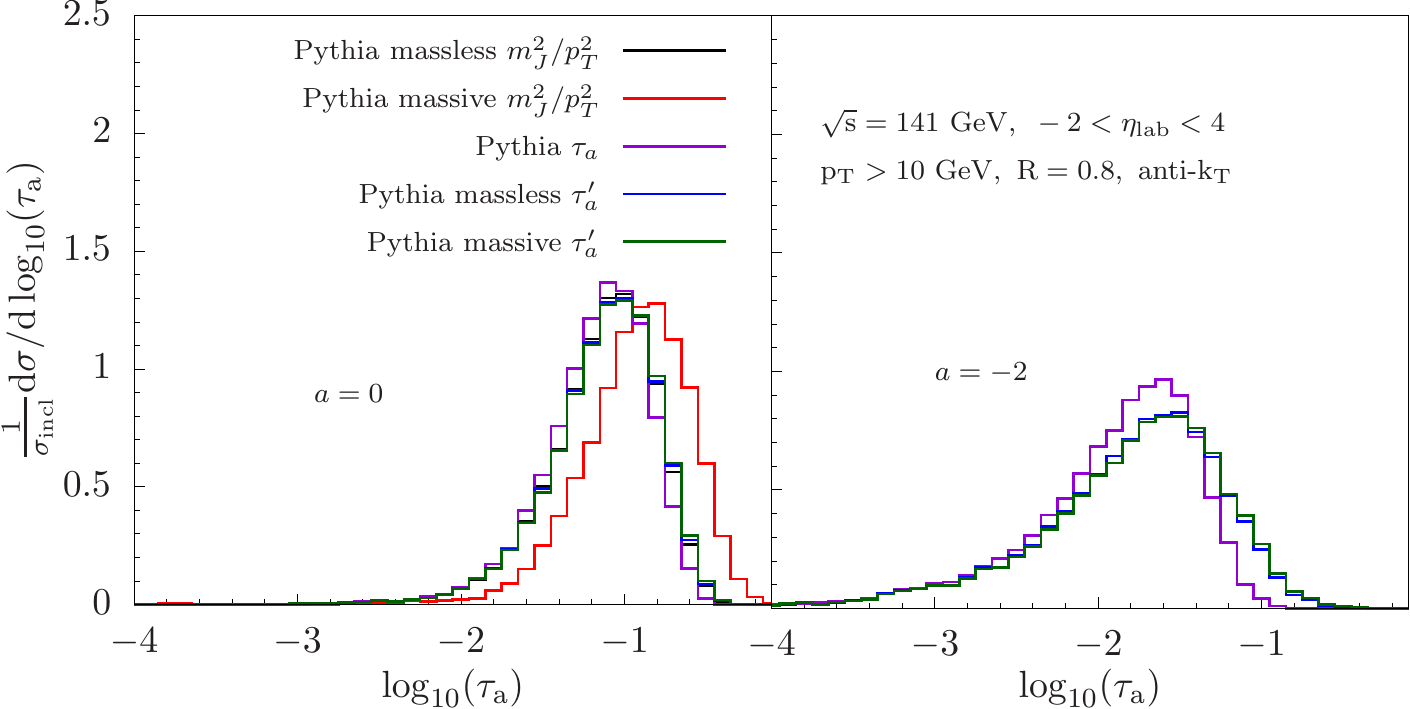}
\caption{Jet angularities in photoproduction at the EIC within the PYTHIA~6 framework for $a=0$ (left) and $a=-2$ (right) using different definitions that agree up to power corrections. See discussion in the text.~\label{fig:a0_am2}}
\end{figure*}

We consider $e+p$ collisions at the EIC at a CM energy of $\sqrt{s} =141~\rm{GeV}$ and a laboratory frame rapidity range of $\-2 < \eta_{\rm{lab}} < 4$. We choose the electron energy as $E_e = 20$~GeV and the proton energy as $E_p = 250$~GeV which corresponds to an expected EIC configuration. The laboratory frame and CM frame 
rapidities are related as
\begin{equation}
\eta_{\rm lab}=\eta+\frac12 \ln\frac{E_p}{E_e} \,.
\end{equation}
In order to estimate QCD scale uncertainties, we vary the scales of the functions appearing in the factorization theorem by a factor of 2 around their canonical values. While varying the individual scales, we maintain the relation
\begin{equation}
\frac{1}{2}\leq \frac{\mu_i}{\mu_i^{\text{can}}} / \frac{\mu_j}{\mu_j^{\text{can}}}\leq 2\,.
\end{equation}
In addition, we also choose to fix the relation between the collinear scale $\mu_C$ and the soft scale $\mu_S$ and for jet angularities also between hard scale $\mu_H$ and jet scale $\mu_{\cal H}$, 
\begin{eqnarray}
\mu_C &=& \mu_S^{\frac{1}{2-a}} (p_T R)^{\frac{1-a}{2-a}} \\
\mu_{\cal H} &=& \mu_H R\,,
\end{eqnarray}
which yields seven different variations. In order to avoid the Landau pole, we smoothly switch off the running of the QCD coupling constant at $450$~MeV using profile scales~\cite{Ligeti:2008ac}.

In~\cite{Jager:2003vy} it was suggested to study the $\eta_{\rm lab}$ distribution which can allow for a separation of the direct and resolved contribution. One of the interesting aspects of studying jets in photoproduction events is that we can gain access to the nonperturbative parton-in-photon PDFs which are poorly understood so far. In particular, the polarized case has never been measured before. Fig.~\ref{fig:jetmul} shows the expected inclusive jet yield at $\sqrt{s}=141~\text{GeV}$ per fb$^{-1}$ of integrated luminosity. Given the expected integrated luminosity of $10$~fb$^{-1}$, we conclude that jets with $p_T \sim 10~\rm{GeV}$ will be produced with sufficiently high statistics at the EIC. We thus study anti-k$_T$ jets~\cite{Cacciari:2008gp} produced with $p_T > 10~\rm{GeV}$ and $Q_{\rm{max}}^2 = 1~\rm{GeV}^2$. An additional cut on the photon momentum fraction (inelasticity) is imposed as $0.2<y<0.8$ due to experimental considerations, see Eq.~(\ref{eq:leptonphoton}). Fig. \ref{fig:incl} shows the comparison of the inclusive jet cross section with $R=0.8$ using the QCD factorization framework and the Monte Carlo parton shower results. We note that here that we vary the hard scale $\mu_H\sim p_T$ and the jet scale $\mu_{\cal H}\sim p_T R$ independently. It is well known in proton-proton collisions that the scale uncertainty of the inclusive jet cross section can be significantly underestimated when both scales are varied together. Several solutions have been proposed in the literature in order to obtain a reliable QCD scale uncertainty estimate~\cite{Dasgupta:2014yra,Currie:2016bfm,Kang:2016mcy,Bellm:2019yyh,Liu:2017pbb}. In this work we vary the two scales independently which can be considered as a very conservative estimate. Overall, we find a good agreement over the entire range of $\eta_{\rm{lab}}$. Note that we use different nonperturbative parton-in-photon PDFs for the resolved contribution for the perturbative (GRS99~\cite{Gluck:1999ub}) and the Monte Carlo calculation (SAS 1D-LO~\cite{Schuler:1995fk}). Nevertheless, we find good agreement since each PDF set has been tuned to similar data sets within the perturbative QCD and Monte Carlo frameworks, respectively. In addition, we note that no hadronization corrections are included in the perturbative results. The remaining hadronization correction is expected to be relatively small for the jet radii and energies considered here and that it is within the conservative uncertainty estimate of our perturbative calculation. It is also worth noting that relative enhancements of the resolved component at large positive $\eta_{\rm lab}$ makes the photoproduction of jets a useful probe for studies of both the unpolarized and polarized parton-in-photon distribution.

\begin{figure*}[t]\centering
\vspace*{.7cm}
\includegraphics[width=5in]{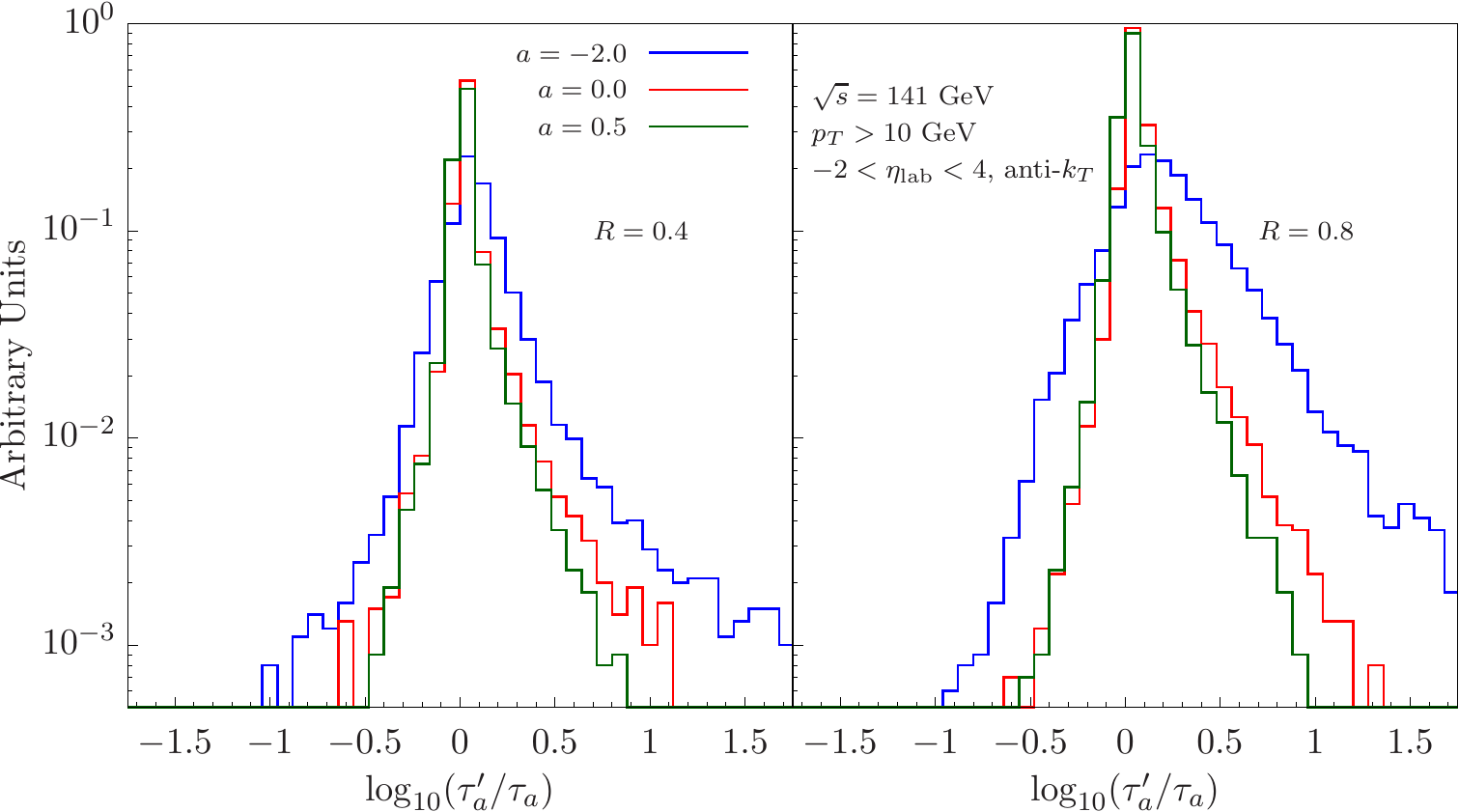}
\caption{Histogram of the number of events plotted as a function of $\log_{10}(\tau_a'/\tau_a)$. We show the results for three different values of $a$ as indicated in the figure for $R=0.4$ (left) and $R=0.8$ jets (right) with the same kinematics as in Fig.~\ref{fig:a0_am2}. Note, all curves have been normalized by a common factor.~\label{fig:pc}}
\end{figure*}

Next, we study jet angularities measured on inclusive jets in photoproduction at the EIC. Using PYTHIA-6 results, we show the distributions for $a=0$ (left) and $a=-2$ (right) in Fig.~\ref{fig:a0_am2} for the same kinematical setup as in Fig.~\ref{fig:incl} above integrated over the rapidity interval of $-2<\eta_{\rm lab}<4$ in the laboratory frame. In order to study the impact of subleading power corrections, we show PYTHIA results using different definitions of the jet angularities, which agree up to power corrections. The purple histogram shows the result using the definition of $\tau_a$ in Eq.~(\ref{eq:taua}) which is given only in terms of transverse momenta and distances in the $\eta$-$\phi$ plane. Second, we show the results using $\tau_a'$ as given in Eq.~(\ref{eq:tauap}) using massless (blue) and massive (green) four-vectors. We include the appropriate prefactor such that the different definitions agree up to power corrections, see Eq.~(\ref{eq:pc1}). We observe only a small difference when massive or massless four-vectors are used which can be understood since mass effects only contribute indirectly to $\tau_a'$ through the jet energy $E_J$ in Eq.~(\ref{eq:tauap}). Third, we consider for $a=0$ also the jet mass definition $\tau_0''$ which is written in terms of a sum over four-vectors squared as given in Eq.~(\ref{eq:tauapp}). The result using massive four-vectors is shown by the red histogram, whereas the black histogram shows the results for massless four-vectors. We observe that for $a=0$ only the red curve deviates significantly from the other curves. This is due to hadron mass effects which directly contribute to the observable when the definition of $\tau_0''$ is used. Therefore, we conclude that it is advantageous to measure jet angularities using the definition of $\tau_a$ or $\tau_a'$ in order to avoid large corrections due to hadron masses. For $a=-2$, we start observing a noticeable discrepancy between the two definitions $\tau_a$ and $\tau_a'$ of the jet angularities indicating that power corrections are numerically more important for smaller values of $a$ which appears to be consistent with the scaling in Eq.~(\ref{eq:NPregion}).

An alternative way to visualize the impact of subleading power corrections is illustrated in Fig.~\ref{fig:pc}. We take all jets produced in PYTHIA that satisfy the selection criteria (same kinematics as in Fig.~\ref{fig:a0_am2}) and calculate the the ratio $\tau_a'/\tau_a$. We plot the number of events that fall in the corresponding bins as a function of $\log_{10}(\tau_a'/\tau_a)$ for $R=0.4$ (left) and $R=0.8$ jets (right). In the case where power corrections are negligible, this distribution should peak at zero. We show the result for three representative values of $a=0.5,\, 0,\, -2$. We observe a reduced yield for $R=0.4$ jets but a narrower distribution indicating a reduced impact of power corrections. This illustrates that $R=0.4$ jets fragment harder whereas $R=0.8$ jets contain more soft particles and can be sensitive to softer scales. Overall, we observe that the distribution is narrower for angularities with larger values of $a$. For $a=-2$, the distribution is significantly broader especially for $R=0.8$ jets which is consistent with the observation made in Fig.~\ref{fig:a0_am2}.

\begin{figure*}[t]\centering
\vspace*{.7cm}
\includegraphics[width=7.1in]{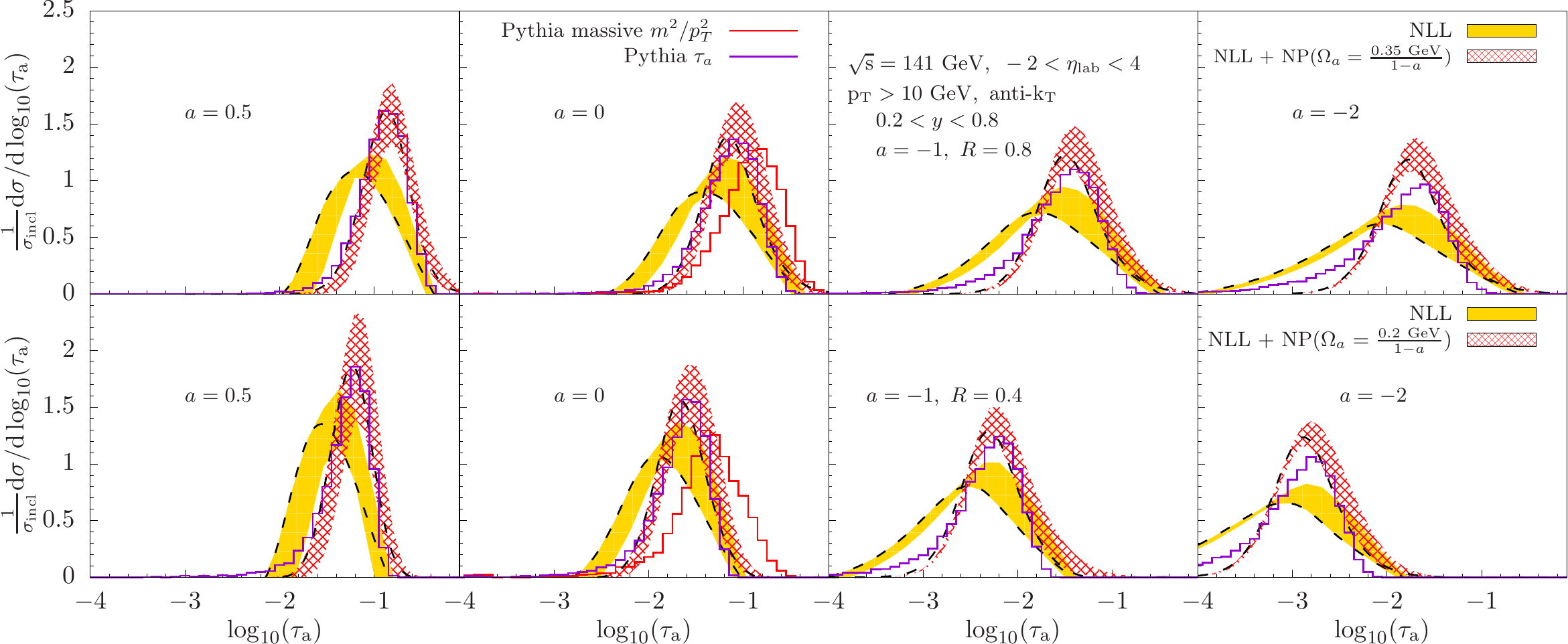}
\caption{Results for jet angularities at the EIC. The purely perturbative results are given by the yellow band, and the results which include nonperturbative effects as introduced via a shape function are shown by the red band. The different panels show the result for different values of the jet radius (top: $R=0.8$, bottom: $R=0.4$) and the parameter $a$ as indicated in the figure.~\label{fig:EIC-NP}}
\end{figure*}

We now compare our PYTHIA~6 results to our perturbative calculations which are shown in Fig.~\ref{fig:EIC-NP}. The purely perturbative results (black dashed line, yellow band) are shown for the same jet kinematics as in Fig.~\ref{fig:incl} and~\ref{fig:a0_am2}. We show the results for five different values of $a=-0.5,\,0,\, -0.5,\, -1,\, -2$ (from left to right) and for two different values of the jet radius $R=0.8$ (upper row) and $R=0.4$ (lower row). The theoretical uncertainties are obtained by varying the scales as discussed above and by taking the envelope. In all panels, we show the Monte Carlo predictions using the definition of $\tau_a$ in Eq.~(\ref{eq:taua}) (purple). In addition, we also show the PYTHIA results using the definition of $\tau_0''$ in Eq.~(\ref{eq:tauapp}) with massive four-vectors (red). All Monte Carlo results here are shown at the hadron level. In order to account for nonperturbative effects, we convolve the purely perturbative result obtained within QCD factorization with a shape function as introduced in Eqs.~(\ref{eq:shapefct1}) and~(\ref{eq:shapefct2}) above. The corresponding result is shown by the dashed black line and red band in Fig.~\ref{fig:EIC-NP}. We find very good agreement for a nonperturbative parameter of $\Omega_{a=0}=0.35$~GeV ($R=0.8$) and $\Omega_{a=0}=0.2$~GeV ($R=0.4$) which is of the order of the expected nonperturbative physics. For $R=0.4$ we need a smaller nonperturbative parameter $\Omega_{a=0}$ which is expected, see the discussion in~\cite{Stewart:2014nna}. We note that the PYTHIA results for $\tau_0''$ shown here in red are outside the uncertainty band of the perturbative results after including the nonperturbative shape function.

We note that the jet angularities $\tau_a$ are generally shifted toward higher values when the jet radius is increased since more particles can be captured in the larger jet. As shown in Eq.~(\ref{eq:NPregion}), the beginning of the nonperturbative region depends on $p_T$ and $R$, which can be identified in Fig.~\ref{fig:EIC-NP} as the region where the QCD scale uncertainty band vanishes. As discussed in the previous section, we need to freeze the running of the strong coupling constant at some value above the Landau pole. In addition, we note that due to the dependence of $\tau_a$  on the distance between the particles in the jet and the jet axis $\sim\Delta R_{iJ}^{2-a}$, the distribution is broader and peaks at lower values for smaller $a$.

\section{Detector considerations \label{sec:detector}}

The pseudo-data results reported in previous sections were obtained at `particle-level', meaning all information for every generated particle was available when constructing jets and calculating angularities. 
Of course, this will not be the case for the actual measurements as particle energies and momenta will be distorted, and some fraction of particles not 
detected at all, due to the finite resolutions and acceptances of any detector. These effects are often evaluated by creating detailed models using 
programs such as GEANT~\cite{Agostinelli:2002hh} that can simulate the response of a detector to an incident particle. Such a detailed study is beyond the scope of this 
article and currently infeasible given that many technical and design choices for an EIC detector have yet to be finalized. However, some evaluation of 
detector requirements can still be made.

Eq.~(\ref{eq:taua}) makes clear that in order to measure angularities, a detector will need to reconstruct accurately both the transverse momenta and 
positions of the produced particles. Regardless of design specifics, there are three primary detector components which will be essential for angularity studies: a tracker, which will 
measure charged particle momenta and trajectories, electromagnetic calorimeters that will measure energies and positions of electromagnetic particles such as 
electrons and photons, and finally, hadron calorimeters which will measure the same for hadrons. Any tracker will provide very good $p_{T}$ and angular 
resolution and the energy and position resolutions of the electromagnetic calorimeters are expected to be good as well. The potentially problematic 
components are the hadron calorimeters, which have poor energy and position resolutions, especially for the low energy particles expected at an 
EIC. Therefore, the remainder of this section focuses on the distortions to angularities induced by the resolution of the hadron calorimeter.

To evaluate the effect that the hadron calorimeters will have on the measured angularity, jets are first reconstructed at particle-level and the angularity is determined. Next, 
the energies and positions of all neutrons and $K_{L}^{0}$s within the jets are smeared by a random amount based on expected resolutions. Energy resolutions 
of $\frac{\sigma_E}{E} = \frac{75\%}{\sqrt{E}} \oplus 15\%$ and $\frac{\sigma_E}{E} = \frac{50\%}{\sqrt{E}} \oplus 10\%$ were taken for mid-rapidity $(\eta_{\rm lab} < 1)$ and forward rapidity $(1 < \eta_{\rm lab} < 4)$, respectively and an overall position resolution of $\sigma_{xy} = \frac{10~\mathrm{cm}}{\sqrt{E}} \oplus 0.6~\mathrm{cm}$ was used. It was assumed that charged hadrons and electromagnetic particles would be detected using the tracker and electromagnetic calorimeters, respectively, with no distortion and only long-lived neutral hadrons would be handled using the hadron calorimeters. With the neutral hadrons 
smeared, the angularity was recalculated and compared to particle-level. 

\begin{figure}[t]\centering
\vspace*{.7cm}
\includegraphics[width=3.3in]{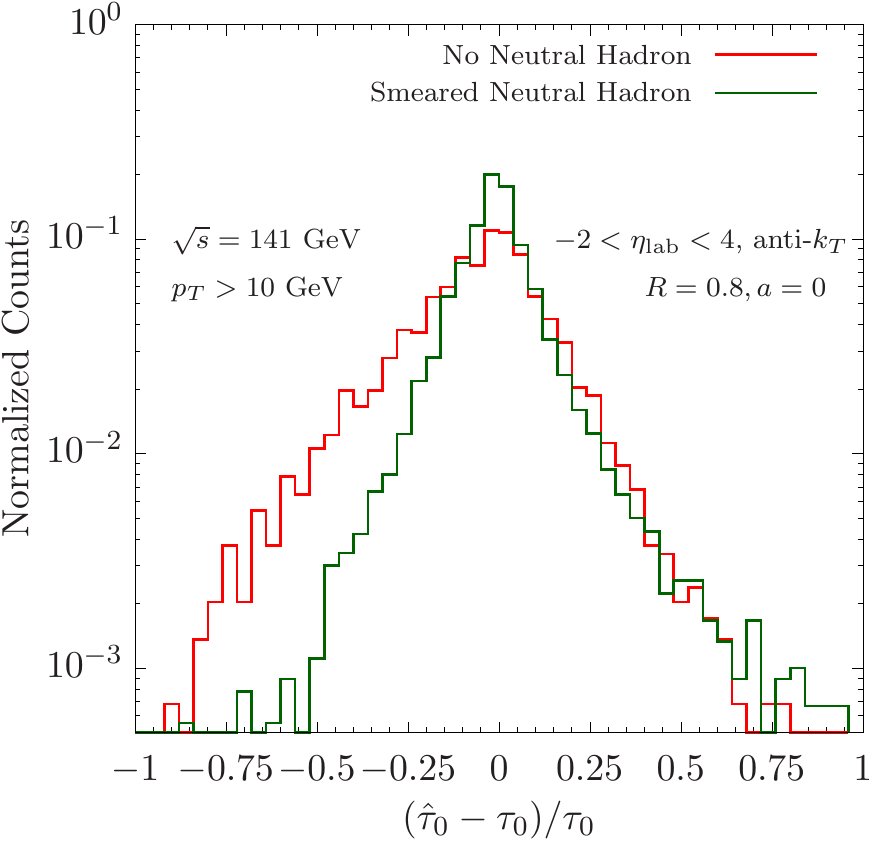}
\caption{Effects of energy and position smearing of neutrons and $K^{0}_{L}$ particles by a hadron calorimeter. Here $\hat{\tau}_{0}$ represents the altered jet angularity while $\tau_{0}$ is the particle level angularity. Each curve is separately normalized to unity and only jets which contained a neutron or $K^{0}_{L}$ are displayed.}
\label{fig:detectorSmearing}
\end{figure}

Fig.~\ref{fig:detectorSmearing} presents the fractional change in angularity due the detector effects described above for the case of $a = 0$ for only 
those jets (roughly 30\% of the total) which contained a neutron or $K_{L}^{0}$. Particle-level jets were required to have $p_{T} > 5$~GeV while the altered 
jet was required to have $p_{T} > 10$~GeV, which allows for jets which smear from below to above the 10~GeV cut to be counted. Also shown is the change which would 
arise in the limiting case that the information from the hadron calorimeters was not used at all and neutral hadrons were not detected. The area of each 
curve is separately normalized to unity so that they can be read as the percentage of jets whose angularity is altered by the amount given on the $x$-axis. 
It is seen that the smeared curve is narrower than the "no-neutrals" curve meaning that the fluctuations 
induced by the assumed calorimeter resolutions are 
less than those which would arise if the hadron calorimeters were not used. If the reverse had been true, it would have meant that such a calorimeter would not have been suitable for angularity measurements. This study does not address whether the overall angularity resolution is suitable for the applications 
listed in the Introduction. That will require a detailed detector simulation and studies of the precision needed to make an impact for each topic and will be 
the focus of future work.

\section{Conclusions and Outlook \label{sec:conclusions}}
%
In this work we performed the first studies of jet substructure observables in electron-proton collisions relevant for the future Electron-Ion Collider. As a representative example, we considered the jet angularity observables, which includes jet mass and jet broadening as special cases. The jet angularity measurements are performed on an inclusive jet sample $e+p\to e' +{\rm jet}+X$ in photoproduction events where jets are reconstructed in the laboratory or center-of-mass frame. The hard perturbative scale of the process is set by the high jet transverse momentum. We performed numerical calculations at NLL$'$ accuracy within perturbative QCD and we calculated the relevant quark/gluon fractions for photoproduction events beyond leading-order using the  NLO code of~\cite{Jager:2003vy}. We compared the perturbative QCD results to our Monte Carlo simulations using PYTHIA~6. The Monte Carlo setup has been tuned to HERA data and we further verified that it reproduces the jet shape data measured by ZEUS/HERA. Hadronization corrections for the perturbatively calculated jet angularity spectrum were included using a suitable nonperturbative shape function. Overall we found good agreement between the two approaches both for the inclusive jet spectrum and the jet angularities. Therefore, our results suggest that jet substructure studies will be feasible at the future Electron-Ion Collider which can complement the current scientific program. In addition, we investigated the numerical size of power corrections within the Monte Carlo setup. By using different definitions of the jet angularities that agree up to power corrections, we investigated their numerical size for the jet angularity spectrum. We found that the corrections are small for jets with transverse momentum $p_T>10$~GeV but they can be sizeable if the jet substructure observable is directly sensitive to hadron masses such as the jet mass when defined as a sum over the four-momenta squared. We conclude that it is important to choose jet substructure observables that are suitable for the relatively low jet transverse momenta and low particle multiplicities that are expected at the Electron-Ion Collider. We also briefly discuss detector requirements needed for accurate experimental measurements of angularity, with a focus on the hadron calorimeter performance. It was found that a hadron calorimeter with energy and position resolutions that could reasonably be achieved in a future EIC detector would contribute useful information to angularity measurements.

The clean environment of electron-proton/nucleus collisions make precision jet substructure studies at the Electron-Ion Collider a unique testing ground of QCD dynamics both in the perturbative and nonperturbative regime. For example, we expect that jet angularities can complement extractions of $\alpha_s$ of DIS event shapes~\cite{Kang:2013nha,Kang:2012zr}. The tuning of parton shower event generators will also greatly benefit from precise jet substructure data in particular when universal nonperturbative components can be determined. Moreover, it will be interesting to use jet substructure observables to investigate cold nuclear matter effects in electron-nucleus collisions.

\section*{Acknowledgments}
%
We are grateful to Miguel Arratia, Barbara Jacak, Peter Jacobs, Zhong-Bo Kang, Yiannis Makris, Mateusz Ploskon, Nobuo Sato, Werner Vogelsang and Feng Yuan for helpful discussions. We would like to thank Werner Vogelsang for providing the NLO code of~\cite{Jager:2003vy}. F.R.\ is supported by the NSF under Grant No. ACI-1550228 within the JETSCAPE Collaboration, by the Department of Energy under Contract No. DE-AC0205CH11231 and the LDRD Program at LBNL.  B.P. and E.C.A. acknowledge the support by the U.S. Department of Energy under contract number DE-SC0012704 and the Program Development Program at BNL.

\bibliographystyle{h-physrev}
\bibliography{bibliography}
\end{document}